\documentclass[acmlarge, screen,authorversion,nonacm]{acmart}

\usepackage{dirtytalk}
\usepackage[framemethod=TikZ]{mdframed}
\usepackage{framed}
\usepackage{xcolor}

\AtBeginDocument{%
  \providecommand\BibTeX{{%
    \normalfont B\kern-0.5em{\scshape i\kern-0.25em b}\kern-0.8em\TeX}}}

\setcopyright{acmcopyright}
\copyrightyear{2021}
\acmYear{2021}
\acmDOI{10.1145/1122445.1122456}

\acmJournal{POMACS}
\acmVolume{37}
\acmNumber{4}
\acmArticle{111}
\acmMonth{9}

\definecolor{shadecolor}{HTML}{D1E8E2}

\mdfdefinestyle{MyShadeQuoteStyle}{
    backgroundcolor=shadecolor,
    linewidth=0pt,
    skipbelow=\topskip,
    skipabove=30pt,
    usetwoside=false,
    innertopmargin=10pt,
    innerbottommargin=10pt,
    innerleftmargin=0.5cm,
    innerrightmargin=0.5cm,
    roundcorner=10pt
}

\begin{document}

\title{Intuitive and Ubiquitous Fever Monitoring Using Smartphones and Smartwatches}

\author{Joseph Breda}
\affiliation{
  \institution{University of Washington}
  \city{Seattle}
  \country{USA}}
\email{joebreda@cs.washington.edu}

\author{Shwetak Patel}
\affiliation{
  \institution{University of Washington}
  \city{Seattle}
  \country{USA}}
\email{shwetak@cs.washington.edu}

\begin{abstract}
Inside all smart devices, such as smartphones or smartwatches, there are thermally sensitive resistors known as thermistors which are used to  monitor the temperature of the device. These thermistors are sensitive to temperature changes near their location on-device. While they are designed to measure the temperature of the device components such as the battery, they can also sense changes in the temperature of the ambient environment or thermal entities in contact with the device. 
We have developed a model to estimate core body temperature from signals sensed by these thermistors during a user interaction in which the user places the capacitive touchscreen of a smart device against a thermal site on their body such as their forehead. During the interaction, the device logs the temperature sensed by the thermistors as well as the raw capacitance seen by the touch screen to capture features describing the rate of heat transfer from the body to the device and device-to-skin contact respectively. These temperature and contact features are then used to model the rate of heat transferred from the user's body to the device and thus core-body temperature of the user for ubiquitous and accessible fever monitoring using only a smart device. We validate this system in a lab environment on a simulated skin-like heat source with a temperature estimate mean absolute error of 0.743$^{\circ}$F (roughly 0.4$^{\circ}$C) and limit of agreement of $\pm2.374^{\circ}$F (roughly 1.3$^{\circ}$C) which is comparable to some off-the-shelf peripheral and tympanic thermometers. We found a Pearson's correlation $R^2$ of 0.837 between ground truth temperature and temperature estimated by our system.
We also deploy this system in an ongoing clinical study on a population of 7 participants in a clinical environment to show the similarity between simulated and clinical trials.
\end{abstract}


\maketitle

\section{Introduction}

Fever (core body temperatures of $38^{\circ}C/100.4^{\circ}$F or higher) is known to be the primary predictive symptom for many viral infections. The most common symptom of the virus behind the 2019 coronavirus pandemic - SARS-CoV-2 - was found to be fever \cite{guan2020clinical}. Similarly, at the end of the 2009 Swine Flu outbreak, the only significant predictive symptoms for positive cases were found to be fever and cough - with fever showing the highest levels of predictability \cite{mahony2013diagnosing}. This finding has also been replicated in a number of studies conducted on influenza \cite{boivin2000predicting, long1997influenza, govaert1998predictive, kohno1995program}. 
This indicates fever as being one of the most important signals for both detection of illness and longitudinal monitor to defend against community spread during an outbreak or pandemic. However, many populations suffer from reduced access to fever monitoring technology due to the thermometer being an expensive and purpose specific device.

From a supply-chain perspective, Improving access to fever monitoring in-home can off-load medical facilities at both the start and end of the treatment pipeline. This is especially important during major outbreaks or pandemics in which healthcare providers can be heavily impacted and tele-medicine, remote screening, and social distancing become essential to prevent the spread of disease while still maintaining adequate healthcare coverage. 

While there exist personal off-the-shelf thermometers, these are purpose specific devices ranging anywhere from ~\$15 to ~\$300 USD. Not only do the price of these devices make them inaccessible to some socioeconomic groups or disadvantaged populations, they can also sell out quickly as seen in the early onset of the 2020 SARS-CoV-2 pandemic, leaving even larger populations without access. On top of this, these devices can be error prone. In a major review of 75 different studies including 8682 patients, peripheral thermometers had pooled 95\% Bland Altman limits of agreement outside the predefined clinically acceptable range of $\pm 0.5^{\circ}C$ (roughly $\pm1^{\circ}F$). Errors were found to be higher in cases of febrile patients with as much as  $-1.44^{\circ}C$ to $1.46^{\circ}C$ (roughly $\pm2.5^{\circ}F$) for adults \cite{niven2015accuracy}. Another study of 25 patients found similarly high Limit of Agreement for tympanic thermometers at $\pm 1.2^{\circ}C$ (roughly $\pm2.2^{\circ}F$) \cite{farnell2005temperature}. This indicates that even existing purpose specific thermometers (both peripheral and tympanic) suffer from inconsistent measurements, bolstering that fever detection is a challenging problem. However, even with large error margins, population scale fever monitoring can still help track the spread of disease. A new route for accessible fever monitoring which can reach all smartphone users with a single software update can have a substantial impact on disease tracking, even if it does not meet the predefined clinically acceptable range. Such a system could be used for remote pre-screening in low resource settings and inform epidemiological models with population scale data. 

A ubiquitous system deployed on existing personal devices could also decrease waste created by the healthcare industry, as most fever screening devices require single-use disposable covers for each test. One study found that if the healthcare sector were a country, it would be the 5th highest emitter, indicating the significant impact these disposables have on the environment.
\footnote{https://noharm-global.org/sites/default/files/documents-files/5961/HealthCaresClimateFootprint\_092319.pdf}.
Therefore, reducing the necessity for purpose specific devices and disposables in the healthcare industry could have a significant impact on the medical industry's climate footprint if scaled - especially in a remote care environment where a unique device is necessary in every home. 

To achieve this ubiquity and access to fever monitoring on existing devices we propose a method of estimating core body temperature using commodity smart devices such as smartphones and smartwatches already present in patients homes. Smartphones have become increasingly ubiquitous in recent years, with nearly 3 billion smartphone users worldwide in 2019, with many of these users in remote or rural areas leveraging smartphones as their primary source of compute power and internet access \cite{taylor2019emerging}. This existing scale and ubiquity lends itself to a number of disease monitoring systems such as contact tracing applications developed as a response to the SARS-CoV-2 pandemic as well as population scale influenza tracking leveraging Wi-Fi connectivity \cite{trivedi2020wifitrace, davalbhakta2020systematic}. Alternatively, smartwatches leveraging heart rate variability tracking have been used to predict SARS-CoV-2 cases as much as a week before diagnosis \cite{warriorWatchStudy}. Adding fever monitoring to these devices could further improve the quality of this public health data and population level disease monitoring systems by expanding the set of features they use with this complementary signal. 

A broad scale deployment of this technology could bring access to population scale fever monitoring in both the developed and developing world and increase the, currently limited, reporting schedule of clinical data to epidemiological models as well as decrease the lag-time of these models \cite{al2020flusense}.

\section{Background}

All smart devices are equipped with a many thermal resistors called thermistors which are used to monitor the health and safety of the smart devices themselves. These thermistors have a variable resistance value which change as a function of the temperature they are exposed to. While these thermistors are designed to monitor the temperature of the device itself, they can sense changes in the temperature of the ambient environment. For example, these thermistors - primarily the thermistor present in the phone battery - have been used in the past to physically model both outdoor air temperature \cite{Overeem2013phonesensor} and ambient indoor air temperature  \cite{breda2019hot, he2020mobile, trivedi29phone}.

In our paper we build on this existing research to leverage these thermistors for sensing core body temperature of a user during an interaction where the device is pressed against the user's body. This new application provides 2 new constraints to the problem: (1) the process of making temperature estimates now becomes an active interaction executed by the user rather than an ambient sensing application, adding user-dependent confounds, and (2) the estimates of temperature are now made over a brief period of time during the active interaction as opposed to longitudinally as in \cite{Overeem2013phonesensor, breda2019hot, he2020mobile, trivedi29phone}. On top of this, we also leverage all available thermistors on the smart device through root access as opposed to the single battery thermistors available to developers through the Android API. These thermistors available through root privileges offer a significantly higher sample-rate than the exposed battery thermistor, making them more desirable for short duration spot estimates. 

We propose that these thermistors - when treated as peripheral thermometers - can provide an accurate proxy measurement of core body temperature. Our approach mirrors the functionality of existing Thermometry by treating the smart device as a temperature probe and leveraging the on-board processing power to predict the steady-state equilibrium temperature based on how heat is transferred from the user's body to the device after a short duration of contact. 
\section{Related Work}

In this section we review the existing literature on using smart devices for temperature sensing and the use of other smart device sensors for mobile health sensing applications.

\subsection{Mobile Health Sensing \& Disease Monitoring Technology}

Recent research in the smartphone health sensing space has shown that there is interest in leveraging these devices for the purpose of sensing bio-signals which can be used for preliminary diagnostic screening. Of these sensing techniques, smartphones have been used to parse rapid diagnostic tests (RDT) \cite{park2020supporting}, monitor blood pressure \cite{Wang:2018:SBP:3173574.3173999}, scan for jaundice \cite{mariakakis2017biliscreen}, and perform spirometry \cite{Larson:2012:SUM:2370216.2370261} all with little to no added hardware by performing sensing using the phone camera or microphone and signal processing directly on the device. While all of these are strong examples of leveraging built-in sensors on smart devices for healthcare purposes, there has yet to be any research in the use of the temperature sensors already present on smart devices to measure core body temperature, despite the strength of such a signal in health diagnostics and disease monitoring. The Huawei Honor Play 4 smartphone released in 2020 does include an infrared temperature sensor which can be used as a thermometer, but this hardware is specific to just this device and takes up valuable real-estate that many phone developers may not consider worth the cost ~\cite{honorPlay4}. Similarly camera based fever monitoring on Android has been done using the FLIR One thermal camera attachment \cite{shinde2020thermotrak}. However, this is a \$400 USD external hardware attachment. Our system uses only hardware present on the Android device out of the box. 

Mobile health technology other than bio-signal sensing has been developed such as leverages mobile phones in aggregate to contact trace throughout a network of individuals and predict community spread. One such example leverages Wi-Fi connectivity of mobile devices to estimate close quarters indoor social interaction
\cite{trivedi2020wifitrace}. Our system could supplement these systems by providing real-time fever data from the population being observed.

\subsection{Passive Temperature Sensing on Smartphones}

In the past, smartphone thermistors have been used to passively measure ambient air temperature of both indoor and outdoor environments  \cite{breda2019hot, he2020mobile, overeem2013crowdsourcing, chau2019estimation, trivedi29phone}. 

One such study modeled ambient air temperature from battery temperature as a function of software usage and physical context of the device. This work leverages a hierarchy of linear and exponential models memorized for the different thermal states of a smartphone to describe heat generated by the device. Activity recognition was used to select the appropriate model which was then treated as a $\Delta T$ component of the battery thermistor signal. This $\Delta T$ was then subtracted from the thermistor signal to reveal ambient air temperature. Evaluation was done by placing many phones in specific states in varying ambient air temperatures and found $1.4\%$ mean error for both spacial and temporal temperature monitoring indoors ~\cite{breda2019hot}. 

A later study developed a similar modeling technique to predict ambient air temperature as the device warms and cools during use. This model also leverages battery current for implicit state detection \cite{he2020mobile}. Evaluation of this model was done inside a thermal chamber which idealizes the evaluation and makes it less generalizable to real rooms. 

Both of these studies specifically modeled the predictable heat generated by phone activity in different usage states and treated the ambient air temperature component of the signal as noise. They then subtracted this modeled temperature from the thermistor signal to reveal ambient air temperature. However, \cite{chau2019estimation} focuses on the thermal behavior of a device in idle state (minimal software utilization) and therefore minimal heat generation on the device. This was previously found to be the software state which most directly reveals ambient air temperature \cite{breda2019hot}. This study then describes the effects of physical contexts such as in-pocket and off-body have on the relationship between an idle phone's temperature and ambient air temperature. 
These passive sensing techniques have also been used for crowd-sourcing both indoor and outdoor temperature. Finally, a nearly decade old study measured temperature at city scale for outdoor environments through the temperature experienced by citizen's devices \cite{overeem2013crowdsourcing}. Similar crowd-sourcing was used by recent work to sense indoor air temperature using a random forest regressorion applied to multiple mobile device's battery temperature signals \cite{trivedi29phone}. 

All of the related work exclusively leverages signals from the battery thermistor. This is the only thermistor which is openly exposed through the Android API. However, our work leverages all 18 thermistors found across the device, located in the $thermal\_zone$ register files. Access to these readings is dependent on root access as only device owner applications (requires root) are allowed to read from the directory containing the $thermal\_zone$ register files. All thermistors across the device exhibit their own unique thermal behavior due to their design and location. Some thermistors experience slower heat transfer likely due to insulating materials inside the device and other thermistors experience slower sample rates or lag in their signal, likely due to hardware or software constraints. 

Another common theme between these studies is the desire to passively measure a longitudinal temperature signal. These all required the system to have a sense of phone state - either due to physical interactions with the device such as where it is stored or software running on the device. Since our work is an active interaction conducted by the user, we are able to control both the software running on the device and the physical context of the device, reducing noise present in our signal by proper design of the study application and user interaction. 
\section{System Design}

In this section we draw analogies between the design of the state-of-the-art clinical thermometer and the hardware present in smart devices to outline the intuition behind our study. We then describe the design of our active sensing system, the physical interactions required to use it, and the data processing involved.

\subsection{Thermometer Technology}

Thermometer technology works by sensing the rate of heat transfer from a user's body to a thermal probe. Higher core body temperatures result in a faster rate of heat transfer. The Welch Allyn SureTempPlus is a state-of-the-art oral thermometer priced on the order of \$300 USD. The thermometer's service manual states that the device consists of a thermistor probe which comes into contact with the patient's body and a predictive algorithm running on the device's microprocessor to predict the temperature of thermal equilibrium that would be reached given enough prolonged contact to reach thermal steady state \footnote{https://www.welchallyn.com}. Similar thermistors are distributed across different components of all smartphones and most wearables. While these thermistors are traditionally used to monitor temperatures of components to ensure safety of the device and user, they can also pick up heat signatures from external sources such as ambient air \cite{breda2019hot} or core body temperature given contact between the device and the user. Drawing an analogy between the thermistors present in both devices allows us to treat the phone as an oversized thermal probe. Using this mental model, the same principals dictating how to use the SureTempPlus can be applied to the smartphone and a similar predictive model can be used to predict core body temperature given prolonged contact between the smart device and the user. However, this new model must account for the noise in the thermal signal due to both the increased size and therefore thermal mass of the smartphone as a probe and the heat generated by software utilization on the device. 

Additionally, the thermistors inside the smart device are embedded at different positions along the device. Intuitively, contact between the user and the device positioned more locally to the thermistor will result in a higher rate of heat transfer from the user to the device. Likewise, a larger surface area of contact between the user and the device will have a similar effect. Heat transfer is also a partial differential equation, meaning that the rate of heat transfer depends on the initial condition of the system. Therefore, a lower starting temperature of the smart device or thermal probe and a higher starting temperature of the user's body will both increase the rate of heat transfer from the user to the phone. 

Our system, much like existing thermometry technology, relies on heat being transferred from the user to the thermal probe. Intuitively, if the temperature of the device or thermal probe is too similar to the temperature of the thermal site on the user's body being measured, there will not be enough heat transfer to make a proper estimate. Further, if the temperature of the device or thermal probe is greater than the user's body, heat will be transferred from the device to the user's body. For simplicity, we only model instances where the user's body temperature is greater than the temperature of the device. 

\subsection{Newton's Law of Cooling}

Equation \ref{equation:newtonsLawOfCooling} shows the physical model describing the idealized thermal behavior of the temperature probe or smartphone as heat is transferred from the user to the device. This equation is idealized and does not take into account the cooling of the warmer thermal body as heat is transferred to the cooler one.

\begin{equation}
    T(t) = (T_0-T_{peak})e^{-kt}+T_{peak}
    \label{equation:newtonsLawOfCooling}
\end{equation}

In this equation $T_0$ represents the initial temperature of the device before contact is made between the device and the user's forehead - for our purposes we use the first thermistor reading at the start of the time series as $T_0$. $T_{peak}$ is the steady-state peak temperature reached by the device after it comes to thermal equilibrium with the user's forehead. $k$ is a parameter for rate of heat transfer which varies between different devices, users, and software running on the device. Ideally, Given enough data this function can simply be fit to the data and $T_{peak}$ would directly reveal the temperature of the user's forehead since the thermal mass of the user is significantly greater than the smart device and the thus thermal equilibrium would occur at roughly the temperature of the user's forehead. However, in practice collecting enough data to curve-fit at test time requires too much time for a user interaction as exponential functions are highly sensitive to noise further from the horizontal asymptote (smaller values of $t$ in Equation \ref{equation:newtonsLawOfCooling}) and curve-fitting is only accurate given a longer duration of contact (15 to 20 minutes).

Prior work used this function and model memorization to model the predictable temperature gain $\Delta T$ of a device due to software activity (for example, the temperature gain due to the screen turning on or the CPU running at various percentages) to make timely estimates of current temperature without curve-fitting at test time. This is possible because each time the device actives a software process it generates a similar amount of heat at a similar rate and thus a model can be memorized for each significant software process \cite{breda2019hot}. However, the temperature gain due to human contact with the device is less consistently predictable than software activity as it varies on confounds such as surface area of contact between the device and the user's body and the distance between the region of contact on the device and the internal location of thermistor. This makes model memorization infeasible as each trial can see a significantly different rate of heat transfer, even on the same user. This requires that a model factor in these variables as features. 

To address this, we use a model to map thermal features captured by the thermistor and screen interaction features captured by the capacitive touch screen to the user's ground truth core body temperature captured by a state-of-the-art oral thermometer. 

\subsection{Approximating Newton's Law of Cooling as Linear}
\label{ssec:Approximating Newton's Law of Cooling as Linear}

\begin{equation}
    T(t) = mt+T_{0}
    \label{equation:linear}
\end{equation}

To make temperature estimates quickly, we approach this problem by collecting only the first 3 minutes of the heat transfer curve shown in Equation \ref{equation:newtonsLawOfCooling} during 3 minutes of contact between a user and the device screen. The exponential curve seen during our trials is notably shallow due to the relatively small difference in the initial temperature of the device and the user's forehead which allows us to approximate the first few minutes of this function with a linear model shown by Equation \ref{equation:linear} rather than trying to fit the true exponential behavior of Equation \ref{equation:newtonsLawOfCooling}. The intuition behind this approximation is that a warmer forehead will transfer more heat to the smart device in the same fixed duration leading to a larger final temperature $T(t)$ resulting in a larger interpolated slope $m$ in our linear approximation. This fixed duration slope is therefore the primary featured used to indicate core body temperature in our model. The other features in our model capture the influence of the user interaction and initial condition of the thermodynamic system on the approximated rate of heat transfer $m$. Our system therefore empirically models a partial differential equation. 

\subsection{Our Smart Device Temperature Sensing System}

\begin{figure}[t]
    \centering
    \includegraphics[width=5in]{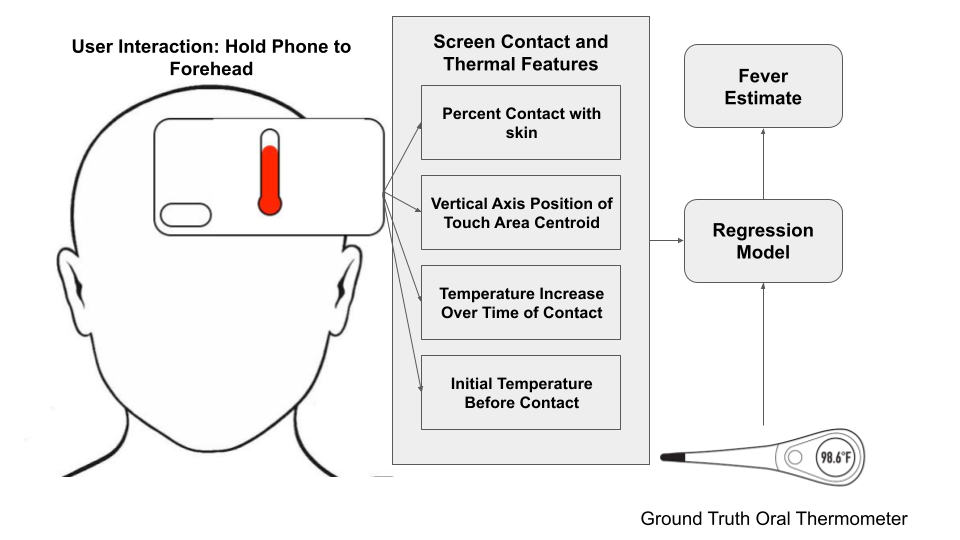}
    \caption{A box diagram showing the user interact and the flow of features into the model to make fever estimates.}
    \label{fig:diagram}
\end{figure}

Unlike previous work in smartphone temperature sensing which have been purely passive, this application is an active sensing technique which requires standardized user interaction to transfer heat from the user to the device to be sensed by the thermistor in a predictable way. Several new challenges and considerations arise from this modification including: (1) a need to define a physical interaction which is intuitive to execute and leads to consistent and repeatable results across users and trials, (2) a brevity constraint to consider a user's attention and respect their time, and (3) awareness of the differences in each trial of this physical interaction between users and executions by the same user which we address both by defining instructions to constrain the user interaction and using the phone touch screen as a sensor to capture features about the interaction. We discuss each of these in turn. Figure \ref{fig:diagram} shows a broad diagram of the interaction as well as a description of the features used in our regression model to make fever estimates.

\subsubsection{Region of Human Body Used to Sense Fever}

Traditionally, core body temperature sensing has been conducted using contact between a thermal probe of a thermometer and a specific thermal site on the body such as the forehead, ear, armpit, or groin. For the purposes of this paper we focused specifically on the forehead as this is an intuitive surface to interact with for most users due to it being well exposed and relatively uniform in shape across users. While other regions of the body such as the armpit can generate even more heat and therefore a stronger signal, they are a less easily accessible thermal site and can vary significantly due to clothing or body shape. For this reason, we constrain our system to predicting core body temperature via contact with the forehead region and leave other thermal sites to future work. 

\subsubsection{Ambient Air Conditions Considerations}

In a 1992 study the forehead temperature most directly correlated to body temperature with some variation as a function of the ambient air. They found with air temperature of $59.0^{\circ}F$ the forehead temperature was around $89.0^{\circ}F$, at $80.6^{\circ}F$ air temperature the forehead temperature was $95.3^{\circ}F$, and at $116.6^{\circ}F$ air temperature the forehead was $98.6^{\circ}F$ which is roughly nominal core body temperature \cite{webb1992temperatures}. This intuitively makes sense as the human body is very good at regulating core-body temperatures, especially in warmer ambient conditions. While this does indicate that the true surface temperature of the forehead may vary as a function of ambient temperature, we chose to simplify our study by conducting all experiments in nominal ambient indoor air conditions of roughly 65-72$^{\circ}F$, resulting in little to no variation in forehead surface temperature due to ambient air temperatures in our study. It is also true that our system relies on the smart device remaining below core-body temperature prior to use as it is built upon the assumption that heat will transfer from the body to the device and not the other way around. This means that in ambient air conditions which would result in a smart device resting at a temperature greater than core-body temperature, the system would not work. This is similarly true for most off-the-shelf thermometers as they are founded on the same principal of heat exchange from the body to the probe. This justifies our choice to design the system for use in temperature controlled indoor environments.  

\subsubsection{Device State}

Similarly, we design our system under the assumption that little heat is generated by the device both before and during the user interaction. We found that in nominal indoor ambient air temperatures of $65^{\circ}F-72^{\circ}F$, idle Android devices with the screen locked and no charger plugged in rest at $70^{\circ}F-74^{\circ}F$. If the device was plugged into a charger this increased the temperature by $1^{\circ}F-2^{\circ}F$ and similarly if the screen was unlocked, the device temperature would increase by $1^{\circ}F-2^{\circ}F$. We found that different thermal behaviors were exhibited while the device was USB charging, AC charging, or the battery was discharging. For simplicity, we conducted all experiments with the device plugged into an AC charger while the device was at $100\%$ charge. This results in the device remaining in a "trickle-charge" state which exhibited a consistent behavior across experiments. We also found that any CPU utilization such as downloads through Wi-Fi or cellular data resulted in dramatic increases in device temperature. For this reason, we left the device in airplane mode for the duration of all experiments. While these constraints were all done manually, a scaled deployment of this system could control the processes and power consumption of the phone during data collection through the operating system to restrict heat generating services while making temperature estimates. In all experiments the device remained below $83^{\circ}F$ at the start of data collection. We aimed to get a balanced distribution of initial condition temperatures for our experiments with temperatures ranging from $70^{\circ}F$ to $83^{\circ}F$ to capture wide array of heat transfer scenarios in our model.

\subsubsection{Grip, Orientation, and Regions of Contact}

\begin{figure}[t]
    \centering
    \includegraphics[width=5in]{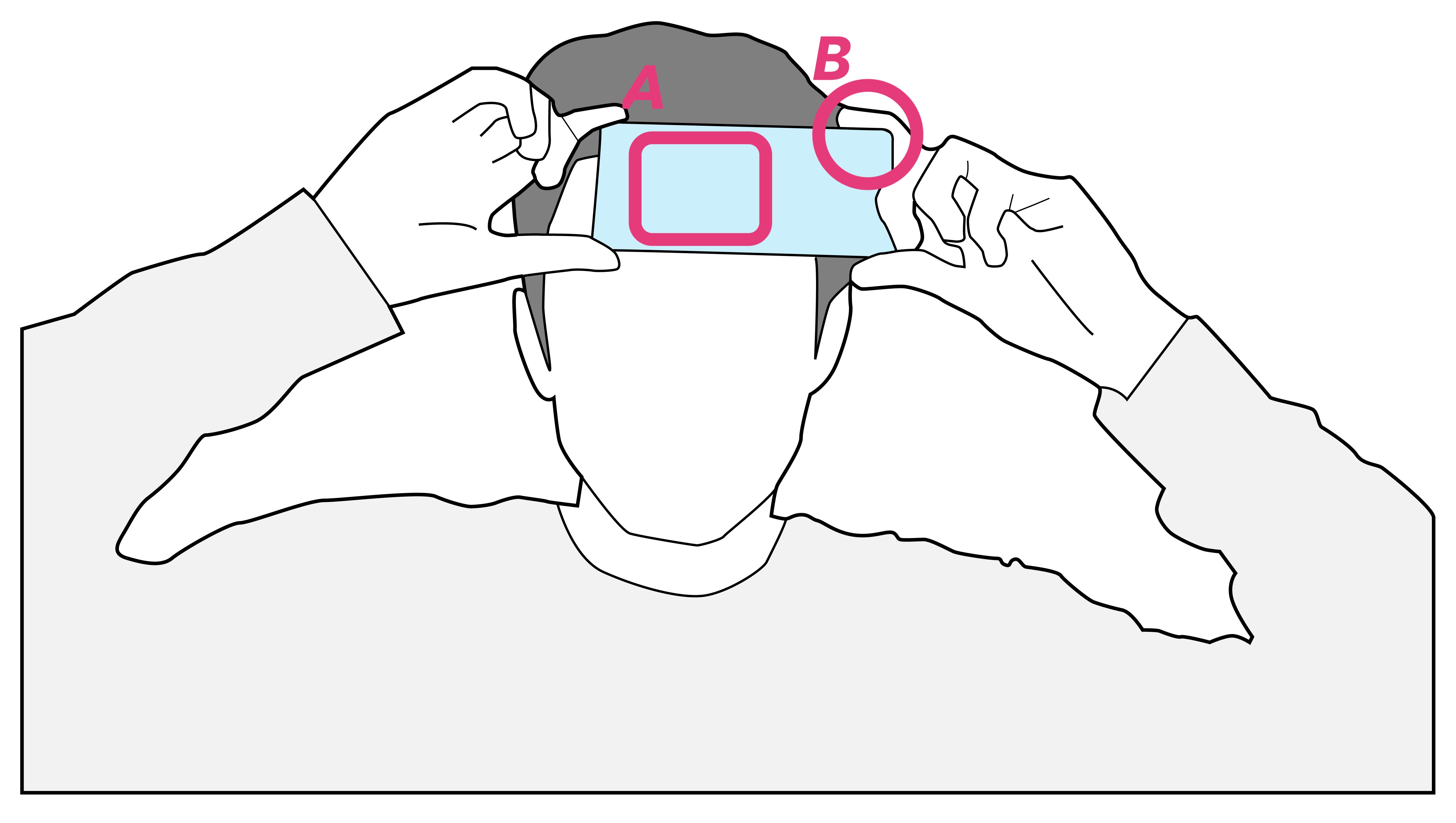}
    \caption{Example posture of user interaction. Box A highlights the region of contact between the user's forehead and device screen closer to the top half of the device, and box B highlights the "camera-grip" style where the four corners of the device are pinched between the user's fingers.}
    \label{fig:phone_posture}
\end{figure}

In order for heat to be transferred from the user to the device in a consistent and repeatable fashion, we defined a standardized interaction to control for grip style, orientation of the device, and region of contact between the user and the device. It is likely that most user's will only be able to bring a portion of the device screen in flush contact with their forehead. It is also likely that the thermistors inside the device are located in different regions of the device, meaning some are more sensitive to contact made in a specific region than others as heat is transferred locally. We found that when the phone is brought to the forehead horizontally as seen in the Box labeled A in Figure \ref{fig:phone_posture}, the surface area of contact spans most of the minor axis of the device. For simplicity, we therefore use the position of the center of mass of contact along the major axis (from the bottom of the device near the microphone to the top of the device near the camera and earpiece) as a feature to describe the location of contact. In preliminary investigation we found that while some thermistors are more sensitive to heat sources located in the bottom of the device, most thermistors are more sensitive to heat sources in contact with the top region of the device. For this reason we instructed all users to aim for contact between their forehead and the top third of the device screen. However, we still collect examples along all locations of the major axis for data coverage in training our model. 

In our preliminary investigation we did not find a significant difference in thermal behavior when holding the device horizontally or vertically against the heat source, so all trials are done with the device held horizontally as it is more convenient for the user. In our preliminary investigation we found that how the device is held by the user can impact the thermal behavior as heat can be exchanged between the user's hands and the device or the ambient air and the device. Since our clinical study is entirely conducted in the same climate controlled facility, we chose to have user's minimize the amount of contact between their hands and the device as we anticipated a higher variance in user's hand temperature than in ambient air temperature between trials. To do this, we developed a "camera-grip" style which can be seen in Figure \ref{fig:phone_posture} in which the four corners of the device are held by the user's index fingers and thumbs in a pinching style similar to how one might grip a point-and-shoot camera. 

\subsubsection{Sensing Amount of Touch on Phone Screen}

We chose to have users orient the device with the screen facing their forehead to allow us to capture capacitance sensed by the touchscreen to describe the amount of contact and location of contact along the device. The Android offers a TouchEvent API to determine location of touch screen interactions. However, this API expects a finger-based interaction and leverages an ellipsoid estimation algorithm to sense touch. This API fails to accurately sense large touch interactions such as contact between a forehead and the device. To work around this, we borrowed a technique from RainCheck \cite{tung2018raincheck}, a system which leveraged a custom kernel to obtain low-level raw capacitance in the form of a 16x32 matrix of capacitance updating every second to distinguish between raindrops and user finger presses. We employed a similar technique, leveraging the same custom kernel to get low-level capacitance from the screen. Conductive surfaces such as human skin or water will increase the capacitance seen by the screen. We used this capacitance matrix to calculate the percentage of the screen in contact with the user by masking out high capacitance values and then calculate the location of contact via the center of mass of this mask along the major axis of the screen. To do this over time, we collected a raw capacitance matrix describing the touch region of the screen every 5 seconds during the trials. We chose to use this low resolution as we wanted to minimize software utilization and therefore any heat generated by the device. We calculated the percent of the screen in contact with the user by averaging all 36 frames of screen capacitance and masking out all cells of the average matrix which had a capacitance greater than a threshold $\tau = 0.25 * C_{max}$ where $C_{max}$ is the maximum capacitance of the averaged matrix and divided the size of this mask by the size of the raw matrix.  

\subsection{Temperature Sensing on Smartwatches}

While the underlying principal of modeling the heat transferred from the user's forehead to the device is the same for both smartphones and smartwatches, the difference in device size and thermistors available change much of the procedure. The reduced size of the device make it likely that most if not all of the device screen will be in contact with the user's forehead. This means we no longer need to track features describing the user's contact with the screen. The watch is also most intuitively pressed against the forehead while being worn on the wrist. We found in preliminary investigation that if the watch was not mounted on the wrist it would launch power management processes which killed our data collection process early. As a result the watch is sandwiched between the wrist and forehead during trials causing it to increase in temperature much faster. The wearable is also significantly smaller and therefore has a smaller thermal mass which also contributes to this quicker rate of heat transfer. Since the temperature increases so quickly, the resulting temperature curve appears much more exponential in the 3 minutes of data collection. As a result, the linear approximation shown in Equation \ref{equation:linear} used for the smartphone is no longer necessary, but may still be applied. But the temperature time series collected from the smartwatch can be fit directly using Equation \ref{equation:newtonsLawOfCooling}. The drawback for using the smartwatch is that it is much more difficult to root and so the only thermistor available to us was the lower resolution battery thermistor. The battery thermistor updates irregularly and based on some internal protocol so sometimes the samples collected would be too sparse (as few as 2 points in 3 minutes) to get a correct fit. We found that given at least 4 points in the 3 minute data collection we could make a fit to get estimates with $0.49^{\circ}F$ Mean Absolute Error (MAE) of core body temperature. This system for temperature sensing on smartwatches is therefore simply curve-fitting the time series retrieved from the thermistor over 3 minutes of contact between the user's forehead and the smartwatch. Given the increased complexity of making estimates using the smartphone, the rest of the paper will focus on the smartphone model unless otherwise noted. 
\section{Lab validation}

\subsection{Experimental Setup}

Gathering data from febrile patients in a clinical setting is challenging because - at the time of writing this - febrile patients are screened out of onsite care due to the SARS-CoV-2 pandemic. Patients who do come onsite with a fever often take anti-inflammatories soon after arriving which can quickly reduce a fever and interfere with data collection. So to validate that the intuition mentioned in Section \ref{ssec:Approximating Newton's Law of Cooling as Linear} resulted in a feature space separable on body temperatures, we simulated a human forehead to the best of our abilities using a zip-lock bag full of water and a sous-vide precision water heater with $\pm1\%$ precision\footnote{The device manufacturer did not expose units of degrees or a baseline temperature that the percent deviation was calculated from.} as specified by the device manufacturer. This provided the opportunity to explore the system in a controlled manner without deviating from the underlying interaction and physics outlined in Section \ref{ssec:Approximating Newton's Law of Cooling as Linear}. 

The sous-vide was used to circulate and maintain water temperature at an array of temperatures between $95^{\circ}F$ and $102.5^{\circ}F$ saturated around the decision boundary of $100.4^{\circ}F$ for a fever. Data was collected for contact between all different locations across the device, but samples were primarily collected around the top third of the device. As mentioned before, data was collected for initial temperatures of the device from $70^{\circ}F$ to $83^{\circ}F$ for good data coverage of possible rates of heat transfer. The sous-vide was positioned inside the water bag consistently in the same location across samples. As mentioned previously, the smartphone was plugged into an AC charger at 100\% charge for all samples to maintain consistent thermal behavior. 

\begin{figure}[t]
    \centering
    \includegraphics[width=4in]{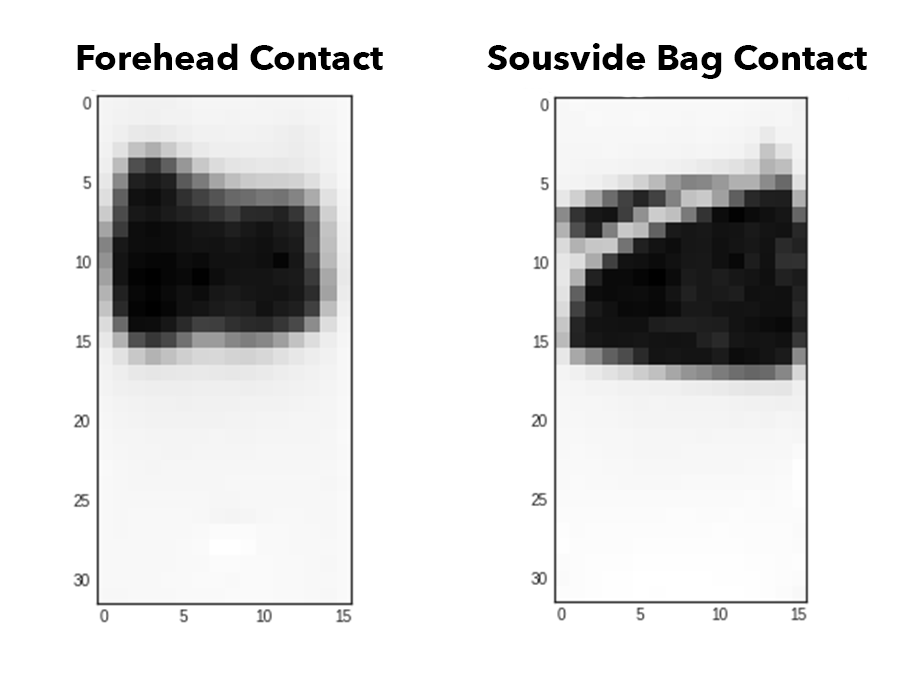}
    \caption{A side-by-side comparison of the contact area of the screen for a human trial and a simulated water bag trial. This demonstrates a consistency between region of the screen and percentage of the screen in contact with the heat source in both trials.}
    \label{fig:contact_compare}
\end{figure}

The physical interaction of holding the device against the forehead was emulated by a cardboard clasp with a length of about one third of the device's length to apply enough pressure to the back of the device to make steady contact between the touch screen and the bag of water. The bag was placed inside a cardboard box with a hole about one third the length of the device cut out to allow only a portion of the screen to maintain contact with the simulated heat source. This was done to control for the region and percent contact made with the device while closely resembling the style of contact made with the device during human trials as seen in Figure \ref{fig:contact_compare} comparing the contact masks of validation and human trials to show the similarity in the screen interaction as detected by the touch screen. 

\subsection{Lab Results: Smartphone}

\begin{figure}[t]
    \centering
    \includegraphics[width=5in]{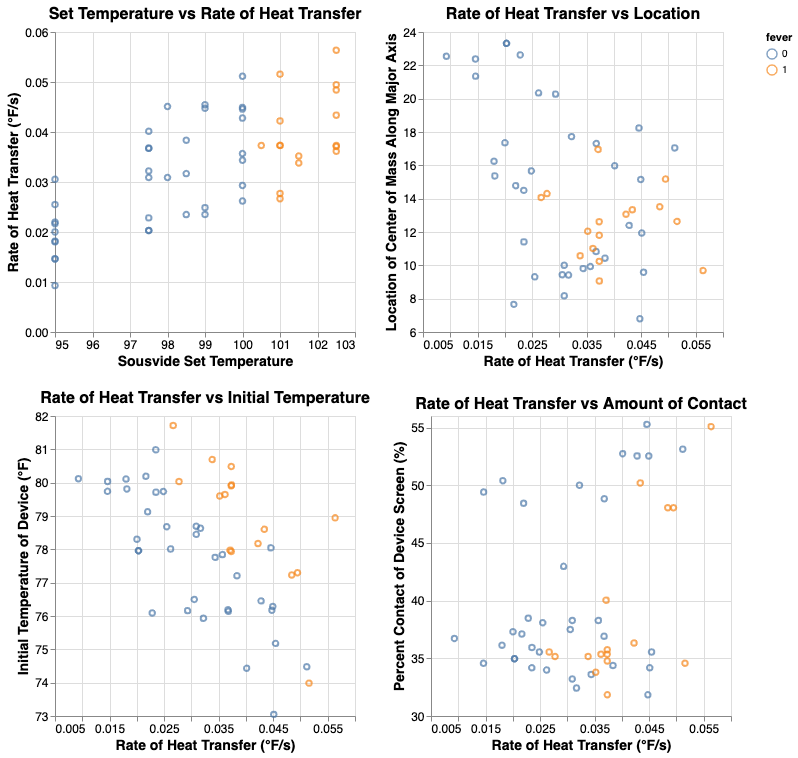}
    \caption{A distribution of all 51 samples from lab validation along the 4 feature axes. Rate of heat transfer is plotted against sous-vide set temperature (top left) to show the distribution of set temperatures captured. Orange points indicate a sample over the $100.4^{\circ}F$ threshold of a fever while blue points represent samples below this threshold. Location of contact (top right), percent of screen in contact (bottom right), and initial temperature of device (bottom left) are plotted against rate of heat transfer for all 3 plots to show each of these features influences the rate of heat transfer.}
    \label{fig:sample_point_distribution}
\end{figure}

Figure \ref{fig:sample_point_distribution} shows the distribution of sample points collected along the 4 feature axes: rate of heat transfer, initial temperature of device at the start of interaction, region of device in contact, and percent of device screen in contact. Each of the 3 features aside from the rate of heat transfer are plotted against the rate of heat transfer to illustrate their respective influence on the rate of heat transfer. Points are colored by the standard fever cut off of $100.4^{\circ}F$. Of the 3 feature spaces, initial temperature of the device plotted against rate of heat transfer is the most visibly separable on simulated fevers. 

\begin{figure}[t]
    \centering
    \includegraphics[width=5in]{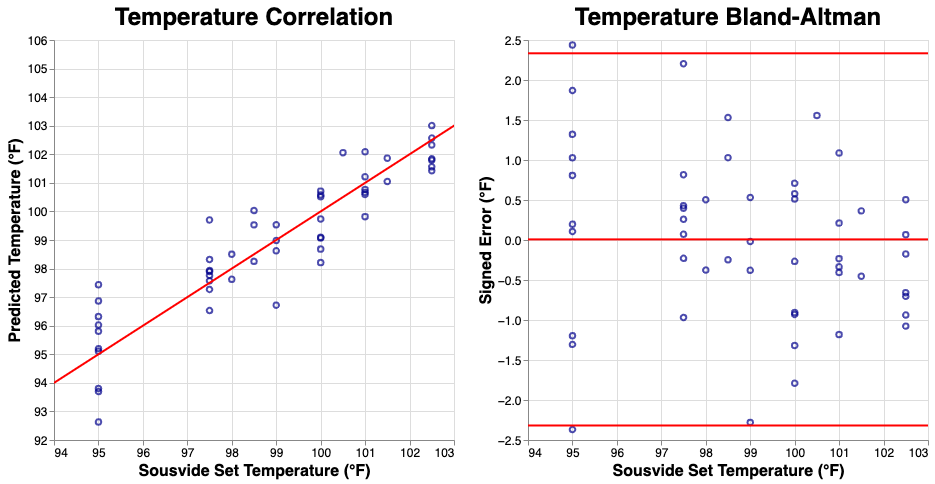}
    \caption{The (left) correlation and (right) Bland-Altman plot for temperature estimates made by a quadratic regression on 51 simulated samples from $95^{\circ}F$ to $102.5^{\circ}F$ controlled by the sous-vide precision water heater. The lines on the Bland-Altman plot (right) show the mean and 95th-percentile limit of agreement.}
    \label{fig:lab_results}
\end{figure}

Different modeling techniques were applied to the simulated data. We found that of Linear Regression, Random Forest Regression, and Quadratic Regression, that Quadratic Regression had consistently the lowest MAE of 0.743 to 0.875 for repeated training, whereas Linear Regression showed an MAE of around 0.953 to 1.107. Random Forest Regression consistently did the worse with upwards of 1.2 MAE. Due to the small sample size of this set, these metrics were calculated by averaging the test-fold error in k-fold cross validation. We tested k=3 through k=10 and found no significant difference in error, but with a small decrease in error for larger k. We found Linear Regression to have a more consistent error across all temperatures, while Quadratic Regression performed better at higher temperatures than for lower temperatures. This is desirable as we would like to ensure detection of fevers in individuals with a temperature above the $100.4^{\circ}F$ threshold. The choice of Quadratic Regression also makes intuitive sense as we are linearly approximating a partial differential equation along multiple non-independent features so we do not expect the decision boundary to be perfectly linear, though we do expect predictable contour. We report Quadratic Regression on the lab validation simulated data to have an MAE of 0.743$^{\circ}F$, and mean error of 0.008$^{\circ}F$ and Standard Deviation of 1.187$^{\circ}F$. Figure \ref{fig:lab_results} shows the correlation and Bland-Altman plots for temperature estimates made by a Quadratic Regression on the simulated data. The model shows an $R^2$ of 0.837 and a Pearson's Correlation between sous-vide set point and predicted temperature of 0.916 with a two-tailed p-value of 4.00e-21. While the regression has a strong fit to the data, the standard deviation 1.187$^{\circ}F$ can partially be contributed to the instability of the sous-vide at maintaining temperature. The sous-vide specifications state a precision of $\pm1\%$ which can be difficult to translate back to concrete temperature precision as units and ranges weren't specified by the manufacturer. We observed that the sous-vide display temperature fluctuated during the 3 minutes of data collection by about $\pm1^{\circ}F$ for some samples. Given a larger sample size with a higher precision and stable heat source, we may have seen a significantly lower standard deviation. 

\begin{figure}[t]
    \centering
    \includegraphics[width=5in]{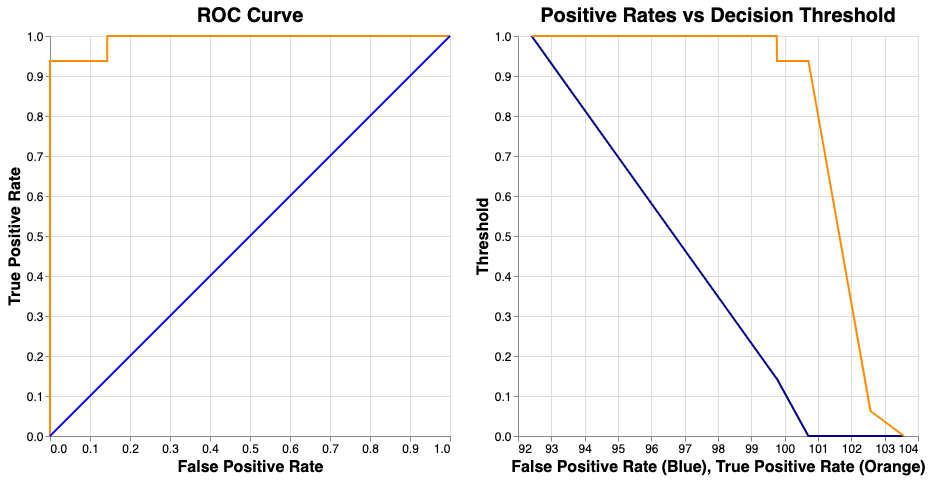}
    \caption{ROC curve (left) showing this system's efficacy as a screening tool for fever monitoring with a 0.991 area under the curve (AUC) and the true and false positive rates as a function of the decision threshold. 100.712$^{\circ}F$ maximizes ROC AUC.}
    \label{fig:roc}
\end{figure}

Figure \ref{fig:roc} shows the receiver operating characteristic (ROC) curve of the Quadratic Regression results after a threshold of 100.712$^{\circ}F$ was applied for binary classification of fever or not. This was implemented using the sci-kit-learn ROC function. We used this method as opposed to directly training a  binary classifier on top of the raw features as this approach is more interpretable than direct binary classification and reflects how one might use a traditional thermometer. That is, a thermometer would measure temperature and provide a reading that is a continuous function of some features sensed by the thermometer and finally a decision would be made on top of that reading to determine if the user has a fever or not. Additionally, we see that the the threshold selected by the sci-kit-learn ROC function was 100.712$^{\circ}F$ which is similar to the original cut off of 100.4$^{\circ}F$ used to binarize the sous-vide set point temperatures as fever or not prior to the regression. This indicates that the regression is slightly over-estimating temperature from the features, but in a way that results in 94\% true positives and 0 false positives if a threshold of 100.712$^{\circ}F$ -- slightly higher than the nominal fever cutoff -- is used. 

\begin{figure}[t]
    \centering
    \includegraphics[width=4.5in]{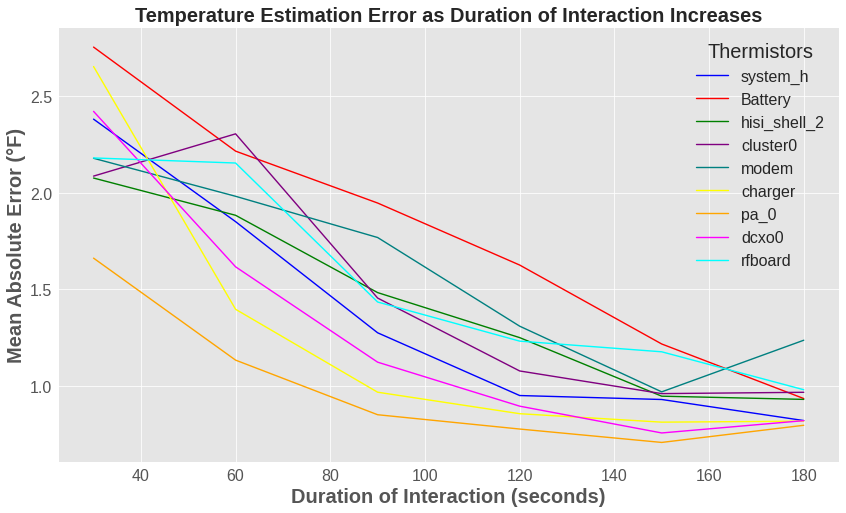}
    \caption{Mean Absolute Error decreasing as duration of data collection increases for a selected subset of thermistors on the device.}
    \label{fig:overtime}
\end{figure}

Figure \ref{fig:overtime} shows the MAE of the quadratic regression for a subset of 9 of the 18 thermistors which performed similarly or better to others at estimating set temperature of the sous-vide as the duration of data collection used in the regression increases in increments of 30 seconds. To get these errors, the regression was retrained using a subset of the 3 minute time series for each duration. Of these 9 thermistors, the battery thermistor performed the worst at shorter durations of data collection. This is due to the low sample rate of this thermistor of around 0.5-1 samples per minute. The thermistors which performed the best were the dcxo0, $pa\_0$, $system\_h$, and charger thermistors, with the $pa\_0$ consistently performing the best with the exception of the charger producing a negligibly better estimate at the 90 second mark. Intuitively, thermistors with higher sample rates will produce better estimates with shorter duration of data collection as their readings more accurately reflect the true slope of the thermal curve. The $pa\_0$ thermistor may have produced the best estimates at lower a duration of contact because it may have a higher sensitivity to external heat sources due to its location on the device or due to its precision and accuracy specifications. 

\subsection{Lab Results: Smartwatch}

\begin{figure}[t]
    \centering
    \includegraphics[width=4.5in]{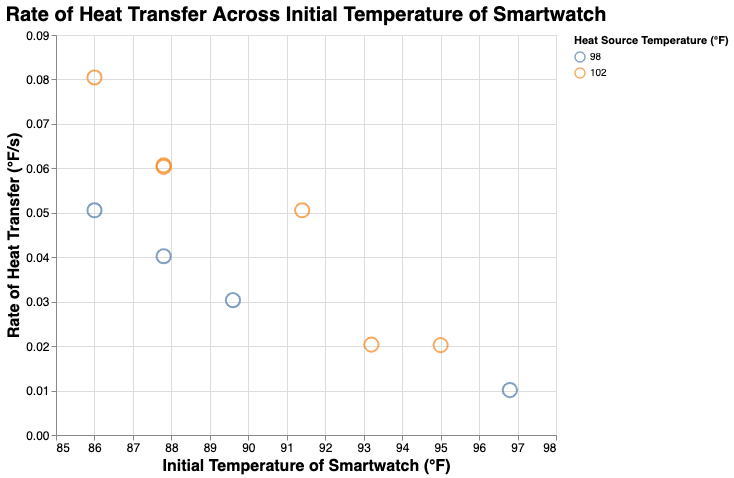}
    \caption{Rate of Heat Transfer from simulated heat source to battery thermistor plotted against the initial temperature of the smartwatch battery before contact.}
    \label{fig:smartwatch_scatter}
\end{figure}

Figure \ref{fig:smartwatch_scatter} Shows the rate of heat transfer - calculated as the last temperature reading minus the initial temperature reading over the total time of contact in seconds - plotted against the initial temperature of the device. Trials were done using the simulated sous-vide heat source at 98$^{\circ}F$ and 102$^{\circ}F$ for simplicity to show the separable behavior between high and low temperatures. Due to the simplicity of the smartwatch scenario, temperature estimates can be made by simply fitting the exponential model to the underlying time series of temperature given a high resolution thermistor, or a similar approach to the smartphone can be taken by leveraging a regression to predict core-body temperature from features extracted from the thermistor time series. However, since the screen interaction no longer varies across trials, a linear model can be used to distinguish between high and low temperatures. 

Figure \ref{fig:smartwatch_scatter} shows that the rate of heat transfer to the smartwatch for high and low temperatures become very similar when the smartwatch temperature is initially close to core-body temperature. This intuitively makes sense as the closer to core-body temperature the device is upon making contact, the less heat will be transferred, resulting in a near-zero rate of heat transfer. For this reason, the smartwatch, like the smartphone, is most effective at estimating core body temperature given a reasonably low steady-state temperature at the time of initial contact. 

\section{Pilot Study in Clinical Setting}

In this section we discuss the pilot preliminary study carried out in a University clinic run by the Department of Family Medicine. The pilot study is still ongoing and is experiencing slow participant recruitment due to febrile participants being screened out of onsite clinical care due to concerns relating to the spread of SARS-CoV-2. However, we include the current results to demonstrate the similarity between simulated and clinical data points as well as address the reasons for dissimilarity. 

We deployed our system with 7 participants. Participants were included on the criteria that they were 18 years or older and visited the clinical location. We considered participants to be febrile if they had a temperature of 100.4 $^{\circ}F$ or higher. Exclusion criteria were patient refusal and inability to give informed consent. Participants who arrived at the clinic would be informed about our study and offered to participate. After obtaining consent, they were given a demonstration of the physical interaction and instructed to hold the device screen firmly and steadily against their forehead for 3 minutes. Every 30 seconds the device would play an audible notification alert which stated the time remaining to maintain participant's attention. After they completed the procedure, their data was uploaded to a database and a unique identifier made to link back to demographic data and ground truth temperature which was collected using an clinical oral thermometer immediately after the interaction. Participants were then offered to repeat the trial again either with the smartphone again or with the smartwatch. If participants were willing to repeat the trial with either device they were instructed to wait at least 10 minutes between trials to allow their forehead to acclimate back to nominal skin temperature and for the smartphone to return to steady state temperature if they were to repeat the smartphone trial.

\begin{figure}[t]
    \centering
    \includegraphics[width=6in]{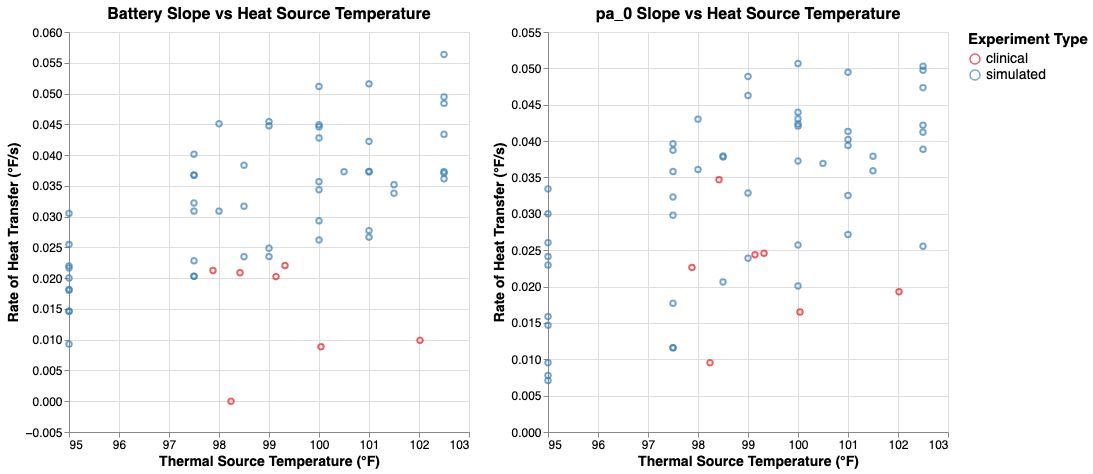}
    \caption{Rate of heat transfer at various thermal source temperatures from both simulated (blue) and clinical (red) trials for both the battery and the $pa_0$ thermistors.}
    \label{fig:clinical}
\end{figure}

Figure \ref{fig:clinical} shows the clinical and simulated data plotted together. It seems that clinical data has a lower rate of heat exchange than the simulated data, but that it may still follow the same trend, just shifted. However, in all clinical trials collected, the percent of the screen in contact with the user was always lower than that of the simulated trials. Clinical trials had anywhere from 8\% to 20\% while the simulated trials had 30\% to 50\% of the screen in contact with the heat source. This is likely due to the participants holding the device more gently against their forehead than the researchers during lab validation. We also found that the rate of heat transfer in simulated and clinical trials were more similar in the $pa_0$ thermistor readings than in the battery thermistor readings. This could be a result of the $pa_0$ thermistor being more sensitive than the battery thermistor and thus picking up more of the external temperature signal given a reduced area of contact.

\section{Conclusion}

This work serves as a foundation for measuring core-body temperature on any smart device with access to its thermistors. The goal of this work is not to replace existing thermometry, but rather to develop a more accessible alternative or proxy in cases where traditional thermometry is not readily available. The systems outlined in this paper are capable of measuring core-body temperature with as little as 0.743$^{\circ}$F (roughly 0.4$^{\circ}$C) MAE and limit of agreement of $\pm2.374^{\circ}$F (roughly 1.3$^{\circ}$C) which is comparable to some off-the-shelf peripheral and tympanic thermometers. We also found a Pearson's correlation $R^2$ of 0.837 between ground truth temperature and temperature estimated by our system. 

While this work is limited in its access to a febrile user population, these results show that there is a strong relationship between the rate of heat transfer to a device and the temperature of the body in contact with it. future work could leverage a much larger clinical study to pursue more complex modeling such as deep learning to map thermistor time series measurements directly to ground truth core body temperature. Future systems could also aim to detect relative change in temperature longitudinally for the same users to track the progression of a fever.

Device manufacturers should consider the utility of the device thermistors discovered in this novel application when designing future board layouts. Future smart devices could be manufactured with the location of thermistors exposed to users, intentionally placed for effective external temperature sensing, and should come packaged with an API for momentarily increasing sample resolution to increase accuracy of spot-sensing temperature. While the rate of heat transfer used for all figures and models in this paper was, by default, calculated over a fixed duration of 3 minutes, we believe that this rate of heat transfer could be obtained through a much shorter interaction given a more precise and higher resolution thermistor. Finally, manufacturers can provide open access data on characteristic thermal curves for different devices which can be derived from logging thermistor data during factory stress tests. Such a dataset could allow for faster model generation of thermal sensing applications by providing a baseline. 

We believe that, if deployed at scale, this system has the potential to improve epidemiology and remote care by allowing healthcare providers to request patients to measure their temperature directly with their smart devices. Patient's data could then be quickly aggregated in real-time at a population scale to decrease the lag-time of epidemiological models and make remote healthcare more continuous and instantaneous for a broad audience. 

\bibliographystyle{ACM-Reference-Format}
\bibliography{bibliography}

\end{document}